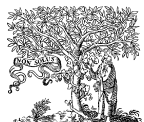

# Photonic density of states maps for design of photonic crystal devices

I.A. Sukhoivanov[a],[*], I.V. Guryev[b], J.A. Andrade Lucio[a], E. Alvarado Mendez[a], M. Trejo-Duran[a], M. Torres-Cisneros[a]

[a]Universidad de Guanajuato, Prol. Tampico 912, Salamanca, GTO 36730, Mexico
[b]National University of Radio Electronics, Lenin Avenue 14, Kharkov 61166, Ukraine

**Abstract**

In this work, it has been investigated whether photonic density of states maps can be applied to the design of photonic crystal-based devices. For this reason, comparison between photonic density of states maps and transmittance maps was carried out. Results of comparison show full correspondence between these characteristics. Photonic density of states maps appear to be preferable for the design of photonic crystal devices, than photonic band gap maps presented earlier and than transmittance maps shown in the paper.
© 2007 Elsevier Ltd. All rights reserved.

## 1. Introduction

At present, photonic crystal (PhC) devices have wide range of applications in telecommunications and lasers as, for instance, PhC fibers [1], laser resonators [2], passive elements, such as high-efficiency waveguides [3], couplers [4], and multiplexers [5]. One of the crucial moments in the design stage of such devices is quick and accurate parameters selection of PhC elements. One of the possible ways to select parameters is the investigation of full photonic band gaps (PBG) maps. PBG maps represent the so-called reduced band structures that give the information about full PBG at different PhC parameters. Such maps were proposed by Joannopoulos et al. [6] and their application for the design of the demultiplexer was shown in [7]. However, such PBG maps provide incomplete information about transmittance characteristics of the PhC because they contain information only about full PBGs and do not indicates the behavior of the PhC outside PBG.

We propose to use photonic density of states (PhDOS) maps instead of full PBG maps analysis. Such PhDOS maps consist of PhDOS taken at different PhC parameters and contain full information about transmittance of finite-size PhC. Peculiar properties of PhDOS of periodic structures were used to modify spontaneous emission of active elements embedded to the periodic structure, and we propose to apply them for the design of PhC devices.

In our work, we have computed the PhDOS map by the plane wave expansion method. Such maps are represented by the PhDOS computed for different radius values of rods. The transmittance spectra of the finite structure at different rod radii for the same structure were also computed. Comparison with PhDOS shows full adequateness of transmittance to PhDOS.

## 2. Photonic density of states computation for 2D PhC

PhDOS is computed based on the band structure of PhC [8]. Photonic band structure can be obtained by the solution of eigenvalue problem for stationary Helmholtz equation given in following form:

$$-\left\{\frac{\partial}{\partial x}\frac{1}{\varepsilon(\vec{r}_{||})}\frac{\partial}{\partial x} + \frac{\partial}{\partial y}\frac{1}{\varepsilon(\vec{r}_{||})}\frac{\partial}{\partial y}\right\} H_z(\vec{r}_{||}) = \frac{\omega^2}{c^2} H_z(\vec{r}_{||}), \qquad (1)$$

where $x$ and $y$ are coordinates in plane of 2D PhC, $\vec{r}_{||}$ is the radius-vector in plane of 2D PhC, $\varepsilon(\vec{r}_{||})$ is the 2D dielectric function of the PhC, $\omega$ is the light frequency, $c$ is the velocity of light in vacuum, and $H_z(\vec{r}_{||})$ is the eigenfunction.

[*]Corresponding author. Tel.: +52 4646880911x221; fax: +52 4646472400.
*E-mail addresses:* i.sukhoivanov@ieee.org, sukhoivanov@salamanca.ugto.mx (I.A. Sukhoivanov).

The most common method for the solution of such problems is the plane waves expansion (PWE) method. It consists of solution of eigenvalue problem for the Helmholtz equation, carried out for an infinite strictly periodic structure by means of representation of wave functions according to Bloch–Floquet theorem, as the product of a plane wave and a periodic function with lattice period:

$$\vec{H}(\vec{r}_{||}) = \vec{H}_{\vec{k}_{||}n}(\vec{r}_{||}) = \vec{H}'_{\vec{k}_{||}n}(\vec{r}_{||})e^{i\vec{k}_{||}\cdot\vec{r}_{||}}, \qquad (2)$$

where $\vec{H}'_{\vec{k}_{||}n}(\vec{r}_{||})$ is a periodic function with lattice period, $\vec{k}_{||}$ is the wave vector, $n$ is the number of eigenstates.

Due to the periodicity, it is convenient to represent the wave function as well as inverted dielectric function as a Fourier series in following form:

$$\vec{H}_{\vec{k}_{||}n}(\vec{r}_{||}) = \sum_{\vec{G}_{||}} H'_{\vec{k}_{||}n}(\vec{G}_{||}) \exp(i(\vec{k}_{||} + \vec{G}_{||}) \cdot r_{||}),$$

$$\frac{1}{\varepsilon(\vec{r}_{||})} = \sum_{G_{||}} \chi(\vec{G}_{||}) \exp(i\vec{G}_{||} \cdot \vec{r}_{||}), \qquad (3)$$

where $\vec{G}_{||}$ is the reciprocal lattice vector in plane of 2D PhC, $\chi(\vec{G}_{||})$ are Fourier expansion coefficients of inverted dielectric function of PhC.

As a consequence of simplification consisting truncation of infinite Fourier series, system of linear equations which is solved numerically is obtained. In case of 2D PhC, the system takes following form [8]:

$$\sum_{\vec{G}_{||}} \chi(\vec{G}_{||} - \vec{G}'_{||})(\vec{k}_{||} + \vec{G}_{||})(\vec{k}_{||} + \vec{G}'_{||})H'_{z,\vec{k}_{||}n}(\vec{G}'_{||})$$

$$= \frac{\omega^{(H)^2}_{\vec{k}_{||}n}}{c^2} H'_{z,\vec{k}_{||}n}(\vec{G}_{||}), \qquad (4)$$

where $H'_{z,\vec{k}_{||}n}(\vec{G}'_{||})$ are wave functions in representation of wave vectors. Searching for eigenvalues of matrix composed of coefficients in wave functions gives us the set of PhC eigenfrequencies for some specific value of wave vector.

In order to obtain the PhDOS, it is necessary to obtain the set of eigenfrequencies for each wave vector value within the first Brillouin zone (see Fig. 1). In case of 2D PhC, the lowest bands of the band structure obtained in this way will be represented by a number of surfaces for one zone each. The band structure represented thus gives full description of spectral properties of PhC and can be used as a basis for the PhDOS computation.

Having the band structure corresponding to the specific type of PhC, the PhDOS $N(\omega)$ is defined by "counting" all allowed states within a given frequency $\omega$, i.e. by the sum of all bands and the integral over the first BZ of a Dirac-$\delta$ function [9]

$$N(\omega) = \sum_n \int_{BZ} d^2k\, \delta(\omega - \omega_n(\vec{k})). \qquad (5)$$

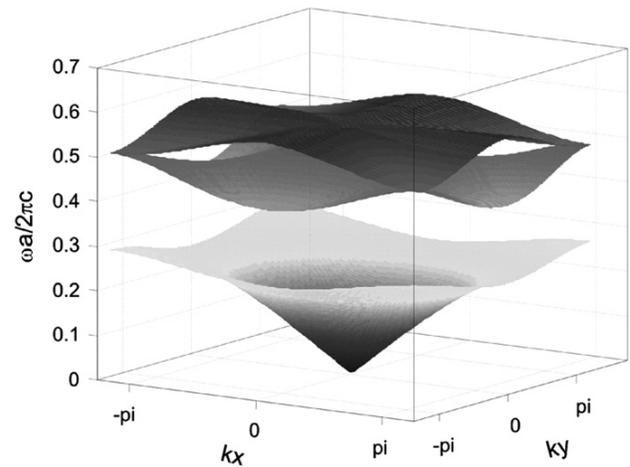

Fig. 1. Band structure of PhC computed for all wave vectors inside 1st Brilluoin zone.

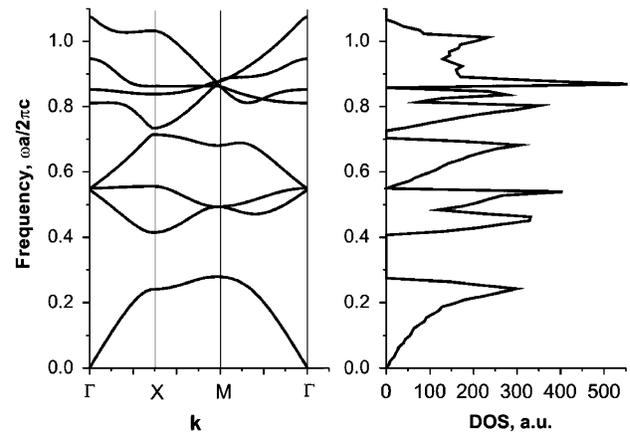

Fig. 2. Band structure and PhDOS of PhC with $n = 3.5$ and $r/a = 0.2$.

In Fig. 2, the band structure computed using the PWE method for the PhC represented by a periodic system of dielectric rods with refractive index $n = 3.5$ and radius $r = 0.2a$ ($a$ is a distance between nearest rods' centers—pitch) separated by air as well as PhDOS of such PhC computed using (5) is depicted. Comparing the obtained characteristics, it can be mentioned that absence or low PhDOS corresponds to full photonic band gap or low number of states in the frequency domain. On the other hand, peaks of PhDOS fall at the domains where plane zones exist or the maximum number of states is concentrated.

Fig. 2 shows the PhDOS for one value of $r/a$ only. However, the solution of device design tasks where it is necessary to determine the PhC parameters from the transmission characteristics requires the information on the PhDOS taken is some range of parameter—PhDOS map. In our work, we computed the PhDOS map for $r/a$ parameter range 0.1–0.4. The example of such PhDOS map is shown in Fig. 3. In the obtained PhDOS diagram, light spots (PhDOS values in range 500–1700) correspond to high PhDOS and dark spots (PhDOS values in range 150–500) correspond to low PhDOS. Black spots (PhDOS values less than 150) correspond to the absence of PhDOS, i.e. full PBG. It can be seen in Fig. 3 that there exist regions where PhDOS values are extremely low. However, they are not equal to zero. This means that they will not be depicted in full PBG maps, as that can cause a mistake while designing the PhC-based device using such maps.

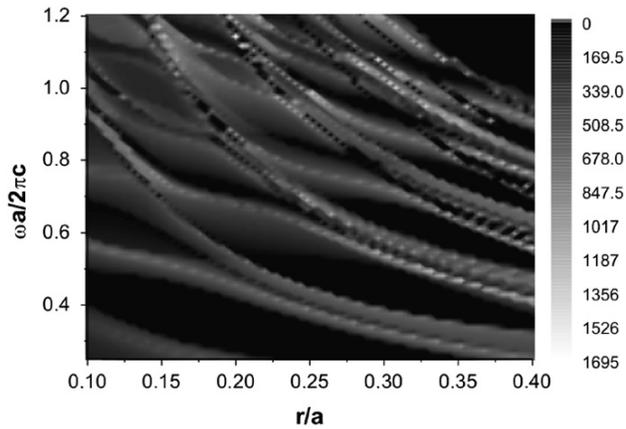

Fig. 3. PhDOS of the PhC made of rods with $n = 3.5$ separated by air.

Thus, the PWE method has been used for the computation of PhDOS. The computation was carried out with the step of $r/a$ equal to 0.002 within the limits of 0.1–0.4. At this, the computation time on the IBM PC with Intel Pentium 2.7 GHz processor amounted 7 h.

## 3. Transmittance maps computation

In order to prove the applicability of PhDOS maps to the design of PhC-based devices, we compared them with transmittance maps of the finite size PhC. In our work, we used the finite differences time domain (FDTD) method for the computation of transmittance maps. The method consists in digitization of Maxwell's equations. At that, differential operators are replaced by differences. For 2D PhC in case of TM polarization [6] the system of Maxwell's equations takes following form:

$$H_{x(i,j)}^{n+1/2} = H_{x(i,j)}^{n-1/2} - \frac{\Delta t}{\mu \Delta y}(E_{z(i,j)}^n - E_{z(i,j-1)}^n),$$

$$H_{y(i,j)}^{n+1/2} = H_{y(i,j)}^{n-1/2} + \frac{\Delta t}{\mu \Delta x}(E_{z(i,j)}^n - E_{z(i-1,j)}^n),$$

$$E_{z(i,j)}^{n+1} = E_{z(i,j)}^n + \frac{\Delta t}{\varepsilon \Delta x}(H_{y(i+1,j)}^{n+1/2} - H_{y(i,j)}^{n+1/2})$$
$$\quad - \frac{\Delta t}{\varepsilon \Delta y}(H_{x(i,j+1)}^{n+1/2} - H_{x(i,j)}^{n+1/2}), \quad (6)$$

where $H_{x(i,j)}^n$, $H_{y(i,j)}^n$ and $E_{z(i,j)}^n$ are field components at nodes of computation mesh, $\Delta t$ is time step, $\Delta x$, $\Delta y$ determine the size of mesh element, $\varepsilon$ and $\mu$ are permittivity and permeability of the medium. The system was solved by assigning initial and boundary conditions. The harmonic signal at one of the boundaries of computation area was used as initial conditions. The cross-section of the signal was a Gaussian function with maximum value in the middle of computation area. Boundary conditions used in the work were represented by the perfectly matched layer (PLM) [10]. The layer is characterized by high value of wave impedance hence high-radiation absorption is emulated. For TM polarization $E_z$ field component is splitted into two parts $E_{zx}$ and $E_{zy}$ for the purpose of introduction of impedance components for different propagation direction. Thus, we obtain the system of four equations in terms of finite differences:

$$H_{x(i,j)}^{n+1/2} = H_{x(i,j)}^{n-1/2} + \frac{\Delta t}{\mu}\left(-\frac{1}{\Delta y}(E_{z(i,j)}^n - E_{z(i,j-1)}^n) - \sigma_y^* H_{x(i,j)}^{n-1/2}\right),$$

$$H_{y(i,j)}^{n+1/2} = H_{y(i,j)}^{n-1/2} + \frac{\Delta t}{\mu}\left(\frac{1}{\Delta x}(E_{z(i,j)}^n - E_{z(i-1,j)}^n) - \sigma_x^* H_{y(i,j)}^{n-1/2}\right),$$

$$E_{zx(i,j)}^{n+1} = E_{zx(i,j)}^n + \frac{\Delta t}{\varepsilon}\left(\frac{1}{\Delta x}(H_{y(i+1,j)}^{n+1/2} - H_{y(i,j)}^{n+1/2}) - \sigma_x E_{zx(i,j)}^n\right),$$

$$E_{zy(i,j)}^{n+1} = E_{zy(i,j)}^n + \frac{\Delta t}{\varepsilon}\left(-\frac{1}{\Delta y}(H_{x(i,j+1)}^{n+1/2} - H_{x(i,j)}^{n+1/2}) - \sigma_y E_{zy(i,j)}^n\right),$$

(7)

where $\sigma_x$ and $\sigma_y$ are x- and y-components of electric field impedance, respectively, $\sigma_x^*$ and $\sigma_y^*$ are x- and y-components of magnetic field impedance, respectively. Computation area and PML regions are shown in Fig. 4. In this work, the graded impedance was used with quadratic dependence on the coordinate to avoid non-physical reflection from the PML [11]. At this, the impedance equals to zero at the inner PML boundary and achieves its maximum value at the outer boundary.

The transmittance spectrum was obtained computing the field distribution at each of the wavelengths within the required range. The transmittance map was obtained by computing the transmittance spectrum for all values of PhC elements parameter in range.

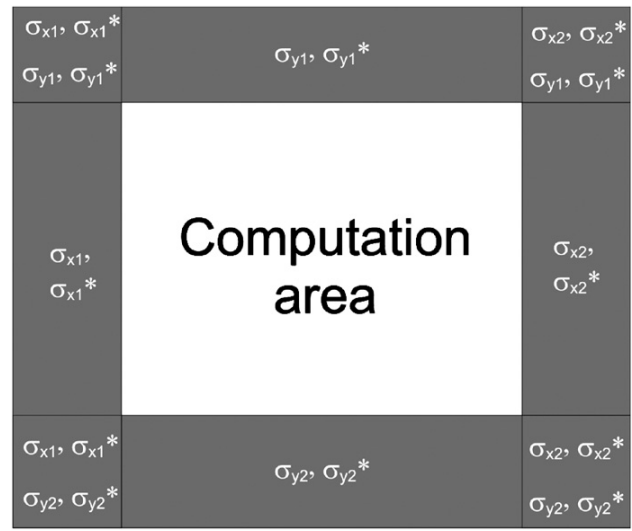

Fig. 4. Computation area for the FDTD method with PML boundary conditions.

The transmittance map computation was carried out for two different $r/a$ steps. In first case, its value was equal to 0.01 and in second case it was 0.002. Moreover, transmittance spectra in first case were computed with 1.5 times less accuracy than in second case. Computation time on the IBM PC with Intel Pentium 2.7 GHz processor amounted 7 h in first case and 120 h in second case. The significant growth of computation time compared to PhDOS maps computation was caused by the fact that computation of transmittance maps by FDTD method requires computation of field distribution at each wavelength at each moment starting from the introduction of radiation to the structure. Hence, increasing the computation accuracy yields significant growth of computation time.

## 4. Results and discussion

Results of transmittance map computation for two cases described above are depicted in Fig. 5. The computations were carried out for the PhC represented by dielectric rods with refractive index $n = 3.5$ separated by air. The range of variation of $r/a$ parameter is 0.1–0.4. At the transmittance map light spots correspond to high transmittance, dark spots correspond to low transmittance and black spots correspond to zero transmittance of the radiation through the finite-size PhC. The comparison between transmittance map (Fig. 5) and PhDOS map (Fig. 3) demonstrates their full correspondence, i.e. high PhDOS (from 500 to 1700 in Fig. 3) corresponds to high transmittance (from 0.5 to 1 in Fig. 5) and full photonic band gap corresponds to absence of transmittance (0 in Fig. 5). From the physical point of view, such correspondence can be explained by the fact that high PhDOS corresponds to high number of wave vectors allowed for the propagation at the specific wavelength in the structure, i.e. high radiation flux through the PhC.

In the work [7], it was shown the application of PBG maps to the design of the 2D PhC-based wavelength division demultiplexer. The design method consists in obtaining geometric parameters of PhC elements forming the input and output waveguide channels as well as PhCs forming band pass filters in the output channels. The scheme of the parameters selection by the analysis of PBG maps is depicted in Fig. 6a. The selection of parameters of the PhC is based on the information about the placement of PBG edges. It is assumed that the transmittance is uniform outside the PBG. At this, in Fig. 6a, $r_1$, $r_2$ and $r_3$ are radii of PhC elements forming the first pass band filter, main waveguide channel and the second pass band filter, respectively. In Fig. 6b it is depicted the PhDOS map with marks pointing operating frequencies (horizontal lines) and radii of PhC elements (vertical ones) both obtained by analysis of PBG map, shown on the Fig. 6a. It can be seen from the PhDOS map that the cross-point of line $r_1$ and line $\omega_1$ lies not in maximum of the PhDOS and, hence, it does not correspond to maximum of transmittance spectrum. Moreover, the cross-point of lines $r_3$ and $\omega_2$ falls at low PhDOS value. Consequently, one of the operating frequencies ($\omega_1$) will not correspond to maximum of transmittance and another one ($\omega_2$) will fall at minimum of transmittance of pass band filter instead of maximum expected from PBG maps.

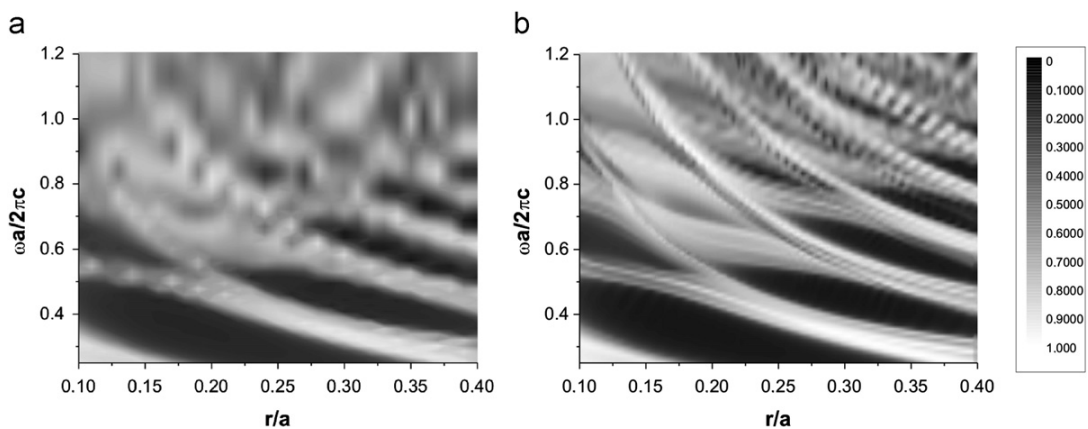

Fig. 5. Transmittance maps of 2D PhC with (a) low resolution and (b) high resolution.

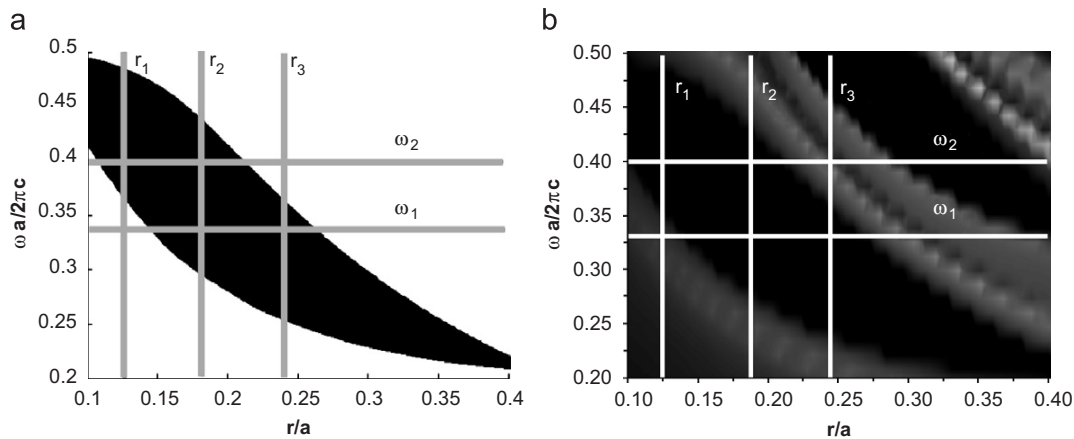

Fig. 6. PBG map (a) and PhDOS map (b) with an example of PhC parameters selection.

Thus, PhDOS maps provide more detailed information about the radiation behavior in periodic structures in contrast to PBG maps because of the capability to describe transmittance of the PhC outside the PBG making them more preferable for the design of the devices on the basis of PhCs.

## 5. Conclusion

In this work, comparison between PhDOS maps and transmittance maps from the point of view of their applicability for designing the PhC-based devices has been carried out. The presented results show full correspondence of the shape of PhDOS and transmittance spectrum. Local extremums of the transmittance spectrum correspond to local extremums of PhDOS in frequency as well as in magnitude. At this, the computation time for the PhDOS is 17 times less than the time required for the computation of the transmittance map.

Thus, utilization of PhDOS maps appears to be more effective for the design of PhC-based devices than the PBG maps because PhDOS maps provide the information about the transmittance of the structure outside the PBG. On the other hand, they are more preferable than transmittance maps because they can be obtained much more quickly but provide the same information about the transmission characteristics of PhC.